\documentclass[a4paper]{jpconf}

\usepackage{graphicx}
\usepackage{latexsym}
\usepackage{amsmath}
\usepackage{pifont}
\usepackage{wasysym}
\usepackage{amsfonts}

\def\be{\begin{equation}}
\def\ee{\end{equation}}
\def\fr{\frac}
\def\ft{\footnote}

\def\pe{\prime}
\def\3s{{s \choose 3}}
\def\4s{{s \choose 4}}
\def\5s{{s \choose 5}}
\def\6s{{s \choose 6}}
\def\12{\frac{1}{2}}
\def\fr{\frac}
\def\pr{\partial}
\def\prd{\partial \cdot}

\def\bec{\begin{center}}
\def\ec{\end{center}}
\def\a{\alpha}  

\def\b{\beta}

\def\d{\delta} 

\def\e{\epsilon}

\def\vf{\varphi}

\def\l{\lambda}
\def\L{\Lambda}
\def\m{\mu}
\def\n{\nu}
\def\r{\rho}

\def\th{\theta}

\def\h{\eta}

\def\cB{{\cal B}}

\def\cE{{\cal E}}
\def\g{\gamma}
\def\cJ{{\cal J}}
\def\cC{{\cal C}}
\def\cL{{\cal L}}
\def\cD{{\cal D}}
\def\cF{{\cal F}}

\def\cA{{\cal A}}
\def\cW{{\cal W}}
\def\cZ{{\cal Z}}

\def\cN{{\cal N}}
\def\cR{{\cal R}}
\def\cS{{\cal S}}

\def\cA{{\cal A}}

\def\dsll{\not {\! \pr}}
\def\psisl{\not {\! \! \psi}}

\def\cWsl{\not {\!\!\! \cal W}}
\def\e{\epsilon}
\def\esl{\not {\! \epsilon}}

\def\xisl{\not {\! \xi}}
\def\esl{\not {\! \epsilon}}

\begin{document}
\title{On the relation between local and geometric Lagrangians for higher spins}

\author{D. Francia}

\address{AstroParticule et Cosmologie (APC)
Universit\'e Paris VII - Campus Paris Rive Gauche
10, rue Alice Domon et L\'eonie Duquet
F-75205 Paris Cedex 13
}

\ead{francia@apc.univ-paris7.fr}

\begin{abstract}
Equations of motion for free higher-spin gauge fields of any symmetry can be formulated in terms of linearised curvatures.
On the other hand, gauge invariance alone does not fix the form of the corresponding actions
which, in addition,  either contain higher derivatives or involve inverse powers of the d'Alembertian operator, thus introducing possible
subtleties in degrees of freedom count. We suggest a path to avoid  ambiguities,  starting from local, unconstrained
Lagrangians previously proposed, and integrating out the auxiliary fields from the functional integral, thus generating
a unique non-local theory expressed in terms of curvatures. 
\end{abstract}

\section{Introduction} \label{sec1}

 One of the unconventional features of higher-spin gauge theories 
 \cite{hsp1, hsp2, hsp3, hsp4, hsp5, hsp6, hsp7, hsp8, hsp9, hsp10} is the peculiar role
played by the corresponding generalised curvatures in the formulation of the dynamics. 

 Higher-spin linearised curvatures  were introduced in \cite{weinberg65, dwf} for symmetric (spinor-) tensors,
as proper extensions of the Maxwell field-strength and of the linearised Riemann curvature\ft{In the so-called
``metric like'' formalism, to be distinguished from the higher-spin generalisation of the Einstein-Cartan formulation 
of General Relativity, usually termed ``frame-like approach'' \cite{hsp1, hsp5}.}. They represent the {\it simplest} gauge invariant tensors 
that  do not vanish on-shell, unless the field itself is pure gauge, with reference to an appropriate generalisation
of the abelian gauge transformations of the photon and of the graviton.
 
 The neat systematics underlying their construction, together with the relevance of their lower-spin counterparts, provide
basic motivations for the idea that they might play some definite role in a theory of higher-spin gauge fields.
On the other hand,  due to the increasing number of derivatives needed in their definition, differently from the cases 
of spin $1$ and spin $2$, for spin $s \geq 3$ it is not possible to derive
from them standard kinetic tensors. For this reason, it was long assumed that one should restrict to the non-geometric 
formulation of Fronsdal \cite{fronsdal}, involving only conventional, second-order,  differential operators for the construction of free equations 
of motion for gauge fields of any spin (for an alternative approach see \cite{sz}). The price to pay with this choice is that,  while no algebraic restrictions (other than symmetry 
properties of the gauge potential itself) are involved  in the definition of higher-spin curvatures, the second-order formulation
of  \cite{fronsdal} requires gauge fields $\vf_{\, \m_1 \, \cdots \, \m_s}$ constrained to be doubly traceless, and subject to an abelian 
gauge transformation involving a rank-$(s - 1)$, traceless gauge parameter:
\be \label{variation}
\d \, \vf_{\, \m_1 \, \cdots \, \m_s} \, = \, 
\pr_{\, \m_1} \, \L_{\, \m_2 \, \cdots \, \m_{s}} \, + \, \cdots \, , 
\ee
\be \label{constraints}
\begin{split}
&\vf^{\, \a \, \b}{}_{\, \a \, \b \, \m_5 \, \cdots \, \m_s} \, \equiv \, 0 \, , \\
&\L^{\, \a}{}_{\, \a \, \m_3 \, \cdots \, \m_{s-1}}  \, \equiv \, 0 \, .
\end{split}
\ee
Under these conditions, it is indeed possible to show that the Fronsdal equation
\be  \label{fronsdalT}
\cF_{\m_1 \, \dots \, \m_s} \, \equiv \, \Box \, \vf_{\m_1 \, \dots \m_s}\, -  \, 
(\pr_{\m_1} \, \pr^{\, \a} \,  \vf_{\a \, \m_2 \, \dots \, \m_s} \, + \, \dots \,) + \, 
(\pr_{\m_1}\pr_{\m_2}\, \vf_{\, \ \a \,\m_3 \, \dots \m_s}^{\, \a}
+ \, \dots \,)\,  = \, 0 \, ,
\ee
propagates the correct polarisations pertaining to a symmetric, massless, spin-s representation
of the Poincar\'e group in $D$-dimensions\ft{To be precise, only tracelessness of the gauge parameter
is needed in the counting of polarisations propagating in \eqref{fronsdalT}. 
Double-tracelessness of $\vf$ is then postulated to construct a suitable gauge-invariant Lagrangian.
See also related discussions in \cite{curtright, dwf}.}. Generalisations of \eqref{fronsdalT}
to the case of tensors of mixed-symmetry were found by Labastida to also involve a proper
set of algebraic conditions, on the traces of gauge fields and of their gauge parameters 
\cite{labastida}. 

 Whereas unusual,  the Fronsdal constraints are instrumental in defining a Lagrangian theory
in which the number of off-shell components is kept to a minimum. Moreover,  the very  
existence \cite{vas90} of consistent, non-linear generalisations of \eqref{fronsdalT}, makes it is fair to say that 
there are no compelling  reasons suggesting that one should try and get rid of them. On the other hand, 
simplifications might be expected in a framework where \eqref{constraints} are not to be assumed 
from the beginning, and in any case it would be rather unsatisfactory if the linearised geometry of 
\cite{weinberg65, dwf} were found not to admit deformations relevant for higher-spin non-abelian 
interactions\ft{Clearly, linearised curvatures can be used to define abelian, 
Born-Infeld type vertices. They also appear naturally  in the quantization of spinning particle models \cite{howe}, as well as in the
description of conformal higher spins \cite{conformalhsp} (for recent results and more references see \cite{conformalnew}).}.

 Indeed, after \cite{dwf}, the study of higher-spin geometry for symmetric tensors was further pursued in \cite{damourdeser}
and then extended to the case of mixed-symmetry tensor fields in \cite{henneauxdb}. Dynamical use of higher-spin curvatures was 
then proposed in \cite{fs1, fs2}, where non-local Lagrangians and equations of motion for symmetric bosons and fermions were 
investigated, with no {\it a priori} reference to the Fronsdal formulation. The latter was  then shown to be recovered performing the same 
partial gauge-fixing required in order to remove all non-localities. In a similar spirit, higher-derivative equations of motion for bosons of mixed-symmetry were formulated in \cite{bb1, dmhull1, bb2} in terms of generalised field-strengths (antecedents of non-Lagrangian
equations formulated via field-strengths can be found in  \cite{weinberg65, sz}) while proposals for corresponding  higher-derivative
or non-local actions were given in \cite{dmhull1, bbhull}. Geometric equations of motion for mixed-symmetry fermions 
were discussed along similar lines  in \cite{hsp4}, while in the symmetric case, quadratic deformations of geometric 
Lagrangians were also constructed \cite{dariomass}, providing direct generalisations of the Proca and Fierz-Pauli theories 
for the description of \emph{massive} fields of any spin.

 The basic indication obtained from these results is that, notwithstanding the presence of higher derivatives or
non-localities, kinetic tensors built out of curvatures can still be used to describe freely propagating
waves of any spin. On the other hand, allowing for non-local operators to be present leads to the consequence that gauge invariance
alone is no more a sufficient criterion for the Lagrangian to be unique, and indeed in \cite{fms1} infinitely many
non-local kinetic tensors were shown to exist, for the case of symmetric bosons, besides the basic ones introduced in \cite{fs1}. As a further selection rule allowing to distinguish among the various options, in \cite{fms1} it was checked whether those theories reproduced 
the correct current exchanges between conserved sources, mediated by massless bosons of spin $s$.
Under this requirement it was found that, out of the infinitely many geometric Lagrangians consistent with 
gauge invariance, \emph{only one} non-local theory  actually possessed  the correct propagator. 

 On the one hand, this result of uniqueness can be considered satisfactory, since it allows 
to define a true candidate linear limit for a theory of interacting higher-spins, possibly
involving non-linearly deformed curvatures. However, it still leaves unanswered a few questions about the meaning
of the whole procedure, given that the check of the propagator is only an \emph{a posteriori} criterion of validity.  
In particular, the very fact that in the absence of sources infinitely many different equations
appear to be consistent, calls for a better understanding of the 
rationale behind the non-localities  of \cite{fs1, fs2, fms1}, 
in the spirit of clarifying  the meaning 
of the manipulations involved. 

 With this purpose in mind, in this note we would like to suggest a different path for the definition of non-local Lagrangians expressed in
terms of curvatures, and compare the outcome with the results of \cite{fs1, fms1}. The idea is very simple:
we consider a framework where Fronsdal constraints are evaded by means of the introduction of auxiliary fields, 
so that the theory is still local, and its physical content can be determined using standard techniques.
In particular, we resort to the Lagrangians proposed in \cite{fs3}, representing the simplest possible unconstrained
ones for the case of symmetric (spinor-) tensors\ft{Previous results leading to non-minimal unconstrained Lagrangians can be found
in \cite{bpt}. The generalisation of the local Lagrangians of \cite{fs3} to the mixed-symmetry case is given in 
\cite{cfms}. Constrained descriptions of mixed-symmetry massless fields in the frame-like approach on maximally
symmetric backgrounds can be found in \cite{mixedframe}.}. 
We then consider the gaussian functional integral for the corresponding theories, and  perform
the integration over the auxiliary fields. In this way we obtain  ``effective'' non-local Lagrangians involving the
physical field $\vf$ alone, whose unconstrained gauge invariance implies that they must be expressible in terms 
of higher-spin curvatures. 

 The advantage of this procedure is twofold: first, we define in this way an \emph{a priori} criterion 
to select one member in the class of all possible non-local, geometric Lagrangians. In addition,
starting from a theory whose spectrum is known by conventional analysis, 
we are  able to unambiguously relate non-localities  to the presence of non-physical fields in the initial local Lagrangians. 

 We investigate along these lines  a few specific cases.
After recalling basic facts about higher-spin curvatures
and minimal local Lagrangians, which we do in Section \ref{sec2}, we study in Section \ref{sec3} the form of the non-local effective action
for symmetric bosons of spin $3$ and spin $4$  on flat space-time. We thus show that the integration
over the auxiliary fields produces  effective, non-local Lagrangians, coinciding with those selected in \cite{fms1}
as the only ones leading to the correct propagators. In addition, in Section \ref{sec3.3} we broaden our analysis to
include the case of (A)dS backgrounds, where we provide the first example of a non-local Lagrangian, for  a spin-$3$ field.
Finally, in Section \ref{sec4} we move our attention to half-integer spins, studying 
the case of fermions of spin $\fr{5}{2}$. We find in this way the form of the  non-local fermionic Lagrangian 
giving rise to the proper current exchange.

 Similar ideas will be exploited in a forthcoming paper \cite{dariotripl} with the purpose of analysing the geometrical content 
of higher-spin \emph{triplets} (\cite{fs2, st, hsp9}, and references therein). In the case of those systems, unconstrained gauge-invariance is related to the
propagation of several irreducible representations\ft{ The unconstrained reduction of triplet Lagrangians to the case of
irreducible spin $s$ is discussed in \cite{triplobuch}. In the frame-like approach, a discussion of the triplets
and of their geometrical meaning can be found in \cite{sorvas}.}, so that, with certain qualifications, a direct correspondence with a sum of
constrained Fronsdal Lagrangians can indeed be established \cite{fottsulcurrent}. Still, once the auxiliary fields are integrated away, the resulting
actions must be expressible in geometrical terms. This suggests in particular that curvatures might play a role in the formulation of higher-spin 
theories, regardless of whether the Fronsdal-Labastida constraints are assumed or not.  
 
\section{Geometric Lagrangians and the issue of uniqueness} \label{sec2}

 Here and in the next section we recall some basic facts, in order to fix the notation and to stress the conceptual issues at stake.

 Following the construction of  \cite{dwf}, for a rank-$3$ tensor subject to the abelian gauge transformation 
\eqref{variation} the corresponding curvature is
\be \label{curvature3}
\cR_{\, \m \m \m, \, \n \n \n} \, = \, \pr^{\, 3}_{\, \m} \, \vf_{\, \n  \n  \n} \, - 
\fr{1}{3} \, \pr^{\, 2}_{\, \m} \, \pr_{\, \n}\, \vf_{\, \m  \n  \n} \, + \,  
\fr{1}{3} \, \pr_{\, \m} \, \pr^{\, 2}_{\, \n}\, \vf_{\, \m  \m  \n} \, - \, 
\pr^{\, 3}_{\, \n} \, \vf_{\, \m \m \m} \, ,
\ee
in a notation where indices denoted with the same letter are to be understood as being completely symmetrised, 
without normalization factors, with the minimum numbers of terms required\ft{ We use the ``mostly-plus'' space-time metric in 
$d$ dimensions, denoted with $\h$.  Apart from the case of curvatures, whenever there is no risk 
of confusion all symmetrised indices are left implicit. Lorentz traces are denoted by ``primes'' or by numbers in square brackets, 
while divergences are denoted by ``$\prd$''. Combinatorial factors can be computed following the rules 
of \cite{fs1, fs2}. In the product of different tensors full symmetrization of indices 
is  always understood, with no weight factors. Useful combinatorial identities are 
\begin{alignat}{4}
\left( \pr^{\, p} \vf  \right)^{\, \pe} \, & =  \, \Box 
\pr^{\, p-2}  \vf  +  2  \pr^{\, p-1}   \prd \vf +  \pr^{\, p} 
\vf^{\, \pe} \, , & \qquad  \ \ \ \  \partial^{\, p}  \partial^{\, q}  \, & = \, {p+q \choose p} 
\partial^{\, p+q} \, , \notag  \\
\left( \eta^k  \vf    \right)^{\, \pe} \, & = \, \left[D
+  2 (s+k-1)   \right] \eta^{\, k-1}  \vf + \eta^k
\vf^{\, \pe} \, , & \qquad \ \ \ \ \eta  \eta^{\, n-1} \, & =   \, n  \eta^{\, n} \, . \notag 
\end{alignat}
}, and where in particular $\pr^{\, k}_{\, \r} =  \underbrace{\pr_{\, \r} \cdots \pr_{\, \r}}_\text{k times}$ is to 
be understood as the product of $k$ gradients.
For spin $4$ the proper generalisation of \eqref{curvature3} is
\be \label{curvature4}
\cR_{\, \m \m \m \m, \, \n \n \n \n} \, = \, \pr^{\, 4}_{\, \m} \, \vf_{\, \n  \n \n \n} \, - 
\fr{1}{4} \, \pr^{\, 3}_{\, \m} \, \pr_{\, \n}\, \vf_{\, \m  \n  \n  \n} \, + \,  
\fr{1}{6} \, \pr^{\, 2}_{\, \m} \, \pr^{\, 2}_{\, \n}\, \vf_{\, \m  \m  \n  \n} \, - \, 
\fr{1}{4} \, \pr_{\, \m} \, \pr^{\, 3}_{\, \n}\, \vf_{\, \m  \m  \m \n} \, + \,  
\pr^{\, 4}_{\, \n} \, \vf_{\, \m \m  \m  \m} \, ,
\ee
while the general formula for spin $s$ in the present notation looks
\be \label{curvatureS}
\cR_{\, \m_s, \,  \n_s} \, 
= \, \sum_{k=0}^{s}\fr{(-1)^{k}}{
\left(
{{s} \atop {k}}
\right)
}\ 
\pr^{s-k}_{\, \m}\pr^{k}_{\, \n} \vf_{\, \m_k,\,\n_{s-k}} \, ,
\ee
with obvious meaning for subscripts.
As previously recalled, the basic property of the tensors \eqref{curvatureS} is their unconstrained
gauge-invariance under the transformation 
\be
\d \vf_{\,  \m_1 \cdots \m_s} = \pr_{\, \m_1}\, \L_{\,  \m_2 \cdots \m_s} + \cdots \, ,
\ee
that in our notation we would simply write as $\d \vf = \pr \L$. 
In addition, they satisfy cyclic and Bianchi identities, reflecting the fact they
define irreducible, two-row Young tableaux for $GL (D, \mathbb{R})$. 
All these properties remain valid if the field $\vf$ carries a spinor index as well, 
so that \eqref{curvatureS} are also suitable for the definition of a linearised fermionic geometry \cite{dwf}.

 In \cite{fms1} the curvature \eqref{curvature3} was used to define  a \emph{one-parameter class} of candidate,
non-local ``Ricci tensors'' for a spin-$3$ field\ft{The subscript ``$\vf$'' in $\cA_{\, \vf}$ is used to distinguish the 
non-local, Ricci-like tensors from their local analogues,  to be introduced in the next section. Those will be indicated with 
the symbol $\cA$, without subscripts, and will depend on the field  $\vf$ and on  an auxiliary field $\a$.}
\be \label{Ricci3}
\cA_{\, \vf} \, (a) \, = \,  \fr{1}{\Box} \, \prd \, \cR^{\, \pe} \, 
+ \, a \, \fr{\pr^{\, 2}}{\Box^{\, 2}} \, \prd \, \cR^{\, \pe \pe}\, ,
\ee 
assumed to define basic equations of motion of the form
\be \label{ricci=0}
\cA_{\, \vf} (a) \, = \, 0 \, .
\ee
A first investigation of the consistency of \eqref{ricci=0} 
led to the observation that, for almost any value of $a$, they can be
shown to imply the ``compensator'' equations
\be \label{compeq1}
\cF \, - \, 3 \, \pr^{\, 3} \, \a_{\, \vf}\, (a) \, = \, 0 \, ,
\ee
where  $\a_{\, \vf}\, (a)$ is a non-local tensor whose explicit form depends on $a$, transforming however 
always with the trace of the gauge parameter:
\be \label{shiftalfa}
\d \, \a_{\, \vf}\, (a) \, = \,  \L^{\, \pe} \, .
\ee
Thus, after a suitable gauge-fixing, infinitely many distinct non-local equations can be reduced to
the Fronsdal form.  In particular, the especially simple choice $a = 0$ reduces \eqref{ricci=0} to the equation
\be
\fr{1}{\Box} \, \prd \cR^{\, \pe} \, = \, 0 \, ,
\ee
a prototype for the results of the works \cite{fs1, fs2}, where free higher-spin Lagrangians 
formulated in terms of curvatures were first proposed.

 On the other hand, as shown in \cite{fms1}, for almost any value of $a$, the actions associated to the 
Ricci tensor $\cA_{\, \vf} (a)$ turn out to give the \emph{wrong} propagator, thus suggesting that the correct counting of degrees of freedom
might involve some further subtleties. 
Moreover, as recalled in the Introduction, \emph{only one} Lagrangian was proven to provide the correct inverse 
kinetic operator, defined by the requirement that the current exchange between distant sources effectively projects 
them onto their transverse-traceless parts in $D-2$ dimensions.
For the spin-$3$ case the  ``correct'' geometric theory is completely characterised by the following quantities:
\be
\begin{split} \label{s3,1}
& \cL \, = \, \12 \, \vf \, \{\cA_{\, \vf} \, - \, \12 \, \h\, \cA^{\, \pe}{}_{\, \vf}\} \, ,\\
& \cA_{\, \vf} \, = \,   \fr{1}{\Box} \, \prd \cR^{\, \pe} \, + \, \fr{\pr^{\, 2}}{2\, \Box^{\, 2}} \, \prd \cR^{\, \pe \pe} 
\, = \, \cF \, - \, 3\, \pr^{\, 3}\,  \a_{\, \vf}  \, , \\
&  \a_{\, \vf} \, = \, \fr{1}{3\, \Box^{\, 2}} \, \prd \, \cF^{\, \pe} \, ,
\end{split}
\ee
where in particular second and third of \eqref{s3,1}  provide the expression of $\cA_{\, \vf} $
in terms of curvatures as well as its compensator form. For spin $4$ the proper Lagrangian and corresponding Ricci tensor  
are instead given by
\be
\begin{split} \label{s4,1}
& \cL \, = \, \12 \, \vf \, \{\cA_{\, \vf} \, - \, \12 \, \h\, \cA^{\, \pe}{}_{\, \vf}\, + \, \h^{\, 2} \, \cB_{\, \vf}\} \, ,\\
& \cA_{\, \vf} \, = \,   \fr{1}{\Box} \, \cR^{\, \pe \pe} \, + \, 
\12 \, \fr{\pr^{\, 2}}{\Box^{\, 2}} \, \cR^{\, \pe \pe \pe} \, - \, 3 \, 
 \fr{\pr^{\, 4}}{\Box^{\, 3}} \, \cR^{\, [4]}
\, = \, \cF \, - \, 3\, \pr^{\, 3}\,  \a_{\, \vf}  \, , \\
& \cB_{\, \vf} \, = \, -\, \fr{3}{8} \, \fr{1}{\Box} \,  \cR^{\, [4]} \, = \, \12 \, 
\{\fr{1}{\Box} \prd \prd\cF^{\, \pe} \, - \, \cF^{\, \pe \pe} \}\, , \\
&  \a_{\, \vf} \, = \, \fr{1}{3\, \Box^{\, 2}} \, \prd \, \cF^{\, \pe} \, - \, 
\fr{1}{3}\, \fr{\pr}{\Box^{\, 3}} \prd \prd \cF^{\, \pe} \, + \, 
\fr{1}{12}\, \fr{\pr}{\Box^{\, 2}}\,\cF^{\, \pe \pe} \, .
\end{split}
\ee
It is clear that the structure of the non-local operators involved in \eqref{s3,1} and \eqref{s4,1} must retain 
some special meaning, quite beyond the fact that they guarantee unconstrained gauge invariance of the 
corresponding Lagrangians, since, as we recalled, the latter can be achieved in several other ways.
As a clue to unconver the rationale behind the  solutions \eqref{s3,1} and \eqref{s4,1} we first recall 
the results of \cite{fs3, fms1}, showing how it is possible to obtain full gauge invariance in the simplest way, 
within the  framework of conventional local theories.

\subsection{Minimal local theory} \label{sec2.1}

For a rank-$s$ fully symmetric tensor the first step is to   
consider the unconstrained variation of the Fronsdal tensor $\cF$,
\be \label{transfgauge}
\d \, \cF \, = \, 3 \, \pr^{\, 3} \, \L^{\, \pe} \, ,
\ee
and introduce  a spin-$(s-3)$ \emph{compensator} field $\a$, transforming as
\be \label{transalfa}
\delta \, \a \, = \, \Lambda^{\, \prime}\, ,
\ee
so that the local kinetic tensor
\be \label{tensorA}
\cA \, = \, {\cF} \, - \, 3 \, \pr^{\, 3} \, \alpha \, ,
\ee
be identically gauge-invariant \cite{fs2, st}.
Then, exploiting the  Bianchi identity for $\cA$,
\be \prd \cA \, - \, \frac{1}{2} \, \pr \, \cA^{\, \prime} \, = \, -
\, \frac{3}{2} \, \pr^{\, 3} \, \{\vf^{\, \prime \prime} \, - \,
4 \, \pr \cdot \a \, - \, \pr \, \a^{\, \prime} \} \, ,
\label{bianchibose} \ee
it is not difficult  to show that a gauge-invariant 
local Lagrangian can be written in the compact form \cite{fs3, fms1}
\be \label{boselagr} 
\cL \, = \, \frac{1}{2} \, \vf \, \{\cA
\, - \, \frac{1}{2} \, \h \, \cA^{\, \pe} \} \, - \,
\frac{3}{4} \ {s \choose 3 } \, \a\, \prd \cA^{\, \pe} \, + \, 3
\, { s \choose 4 } \, \beta \, \{ \vf^{\, \pe \pe} \, - \,
4 \, \prd \a \, - \, \pr \, \a^{\, \pe} \} \, , 
\ee
where the Lagrange multiplier $\beta$ transforms  as
$
\delta \beta \, = \, \prd \prd \prd \Lambda \, ,
$
while the tensor $\cC  \equiv  \vf^{\, \pe \pe} \, - \, 4 \, \prd \a \, - \, \pr \, \a^{\, \pe}$ is the gauge-invariant
completion of the double trace of $\vf$. It is also manifest that Lagrangian \eqref{boselagr} possesses the same 
physical content\ft{In particular, higher derivatives present in the kinetic operator of the compensator $\a$
are harmless. At any rate, it is possible to get rid of them slightly enlarging the field content of \eqref{boselagr} \cite{dariomass, cfms}.} 
as  Fronsdal's one.

\section{Effective non-local Lagrangians for bosons} \label{sec3}

 We would like to establish a direct link between \eqref{boselagr} and  the geometric Lagrangians of \cite{fms1}, 
recalled in the previous section.  The basic observation is that the integration of the auxiliary fields $\a$ and $\b$ 
must define \emph{non-local, effective Lagrangians} for the physical field $\vf$, possessing
the same physical content as  \eqref{boselagr}. The unconstrained gauge invariance of the latter, on the other hand, implies 
that it should be possible to express the resulting effective Lagrangians in terms of curvatures. 

 Here we would like to perform explicitly this computation for the first few cases, to clarify the mechanism
at work in some relatively simple examples. In particular we discuss spin $3$ and spin $4$ on flat space-time, 
and show that the corresponding non-local Lagrangians  actually coincide with those providing the 
correct expression for the current exchange, eqs. \eqref{s3,1} and \eqref{s4,1}. 
In addition, we apply our procedure to the study of a spin-$3$ field on (A)dS background, 
to then pass to the fermionic case, where we concentrate on the example of spin $\fr{5}{2}$ \ft{Since we are going to write functional integrals for gauge theories, we should in principle discuss the issue of gauge fixing, following for instance the Faddeev-Popov procedure. We shall not be concerned with 
this issue here, given that all we want to achieve is an effective Lagrangian for the field $\vf$ alone, whose
proper quantization would require a separate discussion. For similar reasons, we will not discuss the Wick rotation of
our theory to Euclidean space, where in principle functional integrals of gaussian  theories can be given a rigorous 
definition (see e.g. \cite{glimmjaffe}). Rather, we will limit ourselves to the formal use of the rules of gaussian integration in the 
Minkowskian region. This is tantamount to solving the equations for the auxiliary fields and substituting back in $\cL$, 
with some attention to be paid on the field $\b$, for which in the initial Lagrangian there is no full quadratic form.}.

\subsection{Integrating out auxiliary fields: spin $3$ on flat background} \label{sec3.1}

Let us denote with $\cE_{\, \vf}$ the following combination of the  Fronsdal tensor and its trace:
\be
\cE_{\, \vf} \, = \, \cF \, - \, \12 \, \h \, \cF^{\, \pe} \, ;
\ee
the functional integral for the unconstrained spin-$3$ theory, in the presence of an external source, can thus be written
\be \label{Z3}
\cZ \, [\cJ] \, = \, \cN \, \int  \cD \vf \, \cD \a \, 
e^{\, i \, \int  d^d x \{ \12 \, \vf \, \cE_{\, \vf} \, + \, \fr{9}{4} \, \a \, \Box^2 \, \a \, 
- \, \fr{3}{2}\, \a \, \prd \cF^{\, \pe} \, - \, \vf \, \cdot \, \cJ\}}\, ,
\ee
where $\cN$ is an overall normalization. Performing the integration over $\a$ one obtains
\be \label{Z3}
\begin{split}
\cZ \, [\cJ] \, &= \, \cN_{\, \a} \, \int  \cD \vf \,  
e^{\, i \, \int  d^d x \, \{\cL_{\, eff} \, (\vf) \, - \, \vf \, \cdot \, \cJ\}} \, \\
&=  \, \cN_{\, \a} \, \int  \cD \vf \,  
e^{\, i \, \int  d^d x \{ \12 \, \vf \, \cE_{\, \vf} \, - \, \fr{1}{4} \, \prd \cF^{\, \pe} \, \fr{1}{\Box^2} \,  \prd \cF^{\, \pe}  \, 
 - \, \vf \, \cdot \, \cJ\}}\, ,
\end{split}
\ee
where $\cN_{\, \a}$ indicates a properly modified, field-independent, normalization factor, taking into account  the gaussian
integration over $\a$.
The corresponding  non-local, effective Lagrangian is
\be \label{Leff3}
\cL_{\, eff} \, (\vf) \, = \, \12 \, \vf \, \cE_{\, \vf} \, - \, \fr{1}{4} \, \prd \cF^{\, \pe} \, \fr{1}{\Box^2} \,  \prd \cF^{\, \pe} \, ,
\ee 
and can be shown to coincide, up to total derivatives, with the one defined in  (\ref{s3,1}).

\subsection{Integrating out auxiliary fields: spin $4$ on flat background} \label{sec3.2}

For the unconstrained spin-$4$ case, rearranging terms in \eqref{boselagr}, we can write the functional
integral as
\be \label{Z4}
\cZ \, [\cJ] \, = \, \cN \, \int  \cD \vf \, \cD \b \, \cD \a \, 
e^{\, i \, \int  d^d x \{ \12 \, \vf \, \cE_{\, \vf} \, + \, 3\, \b \, \vf^{\, \pe \pe} \, 
+ \, \12 \a \, A_{\, \a} \, \a \, + \, \a \, B_{\, \a} (\vf, \, \b) \, - \, \vf \, \cdot \, \cJ\}}\, ,
\ee
where
\be \label{ABoper}
\begin{split}
A_{\, \a} \, & = \, 18 \, \Box \, (\Box \, + \, 3 \, \pr \, \prd) \, , \\ 
B_{\, \a}  \,& = \, - \, 6 \, \{\prd \, \cF^{\, \pe} \, + \, \pr \, \prd \prd \vf^{\, \pe} \, 
- \, \Box \, \pr \, \vf^{\, \pe \pe} \, - \, 4 \, \pr \, \b\} \, ,
\end{split}
\ee
are the relevant operators  entering the gaussian integration over $\a$.

Technically, the main issue consists in the inversion of the operator
$A_{\, \a}$, whose form depends crucially on the rank of the tensors 
on which it is supposed to act. In the present case what we need
is  $A^{\, -1}_{\, \a}$ when acting on the space of rank-$1$ tensors (the rank of 
the compensator $\a$ associated to the rank-$4$ field $\vf$), whose explicit expression is
\be \label{invAspin4}
A^{\, -1}_{\, \a} \, = \, \fr{1}{18 \, \Box^{\, 2}} \, (1 \, - \, \fr{3}{4} \, \fr{\pr}{\Box}\, \prd ) \, .
\ee
Performing the integration over $\a$ in \eqref{Z4}, and using \eqref{invAspin4}, we thus obtain
\be \label{Z4alfa}
\cZ \, [\cJ] \, = \, \cN_{\, \a} \, \int  \cD \vf \, \cD \b \, 
e^{\, i \, \int  d^d x \{ \12 \, \vf \, \cE_{\, \vf} \, + \, 3\, \b \, \vf^{\, \pe \pe} \, 
+ B_{\, \a} (\vf, \, \b)\, \fr{1}{18 \, \Box^{\, 2}} \, (1 \, - \, \fr{3}{4} \, \fr{\pr}{\Box}\, \prd )\, 
B_{\, \a} (\vf, \, \b)\,  - \, \vf \, \cdot \, \cJ\}}\, .
\ee
In the latter expression we can reorganize the various terms, so as to make it explicit the quadratic form in the field
$\b$, allowing to perform the second gaussian integration:
\be\label{Z4alfa2}
\cZ \, [\cJ] \, = \, \cN_{\, \a} \, \int  \cD \vf \, \cD \b \, 
e^{\, i \, \int  d^d x \{ \12 \, \vf \, \cE_{\, \vf} \, - \, C_{\, \vf} \, 
+\, \b \, \fr{1}{\Box} \, \b \, + \, \b \, (3 \, \vf^{\, \pe\pe}\, - \, \fr{1}{\Box^{\, 2}} \, \prd D_{\, \vf} )\, 
  - \, \vf \, \cdot \, \cJ\}}\, ,
\ee
where
\be \label{CDoper}
\begin{split}
C_{\, \vf} \, & = \, D_{\, \vf} \, \fr{1}{\Box^{\, 2}} \, D_{\, \vf} \, 
+\, \fr{3}{4} \, \prd D_{\, \vf} \, \fr{1}{\Box^{\, 3}} \, \prd D_{\, \vf} \, , \\ 
D_{\, \vf}  \,& = \, \prd \cF^{\, \pe} \, + \, \pr \, \prd \prd \vf^{\, \pe} \, - \, \12 \, \Box \, \pr \, \vf^{\, \pe\pe} \, .
\end{split}
\ee
Integrating over $\b$ we are formally left with a  theory involving the field $\vf$ alone
\be 
\cZ \, [\cJ] \, = \, \cN_{\, \a, \, \b} \, \int  \cD \vf  \, 
e^{\, i \, \int  d^d x \{ \cL_{\, eff}\, (\vf)\, 
  - \, \vf \, \cdot \, \cJ\}}\, ,
\ee
where again the resulting non-local, effective Lagrangian
\be \label{effectivespin4}
\cL_{\, eff}\, (\vf) \, = \, \12 \, \vf \, \cE_{\, \vf} \, - \, D_{\, \vf} \, \fr{1}{\Box^{\, 2}} \, D_{\, \vf} \, 
- \, \fr{3}{4} \, \prd D_{\, \vf} \, \fr{1}{\Box^{\, 3}} \, \prd D_{\, \vf} \, -\, \fr{3}{2} \, \vf^{\, \pe \pe} \, 
\fr{1}{\Box} \, \prd D_{\, \vf} \, + \, \fr{9}{4} \, \vf^{\, \pe \pe} \, \Box \, \vf^{\, \pe \pe} \, ,
\ee
can be shown to be equivalent to the corresponding expression in \eqref{s4,1}, 
for instance computing the equations of motion of \eqref{effectivespin4} and verifying that they can be
rearranged in the form
\be
\cA_{\, \vf} \, - \, \12 \, \h\, \cA^{\, \pe}{}_{\, \vf}\, + \, \h^{\, 2} \, \cB_{\, \vf} \, = \, 0 \, ,
\ee
with the various quantities in this expression defined in \eqref{s4,1}.

\subsection{Integrating out auxiliary fields: spin $3$ on (A)dS background} \label{sec3.3}

 In \cite{fms1} the minimal local Lagrangians \eqref{boselagr}, together with their fermionic counterparts, 
to be recalled in Section \ref{sec4}, were also generalised to the case of (A)dS backgrounds. On the other hand, 
the analysis of the non-local theory was considered only for the bosonic case on flat space-time. Here we would like to perform a first step
towards a more general discussion, describing the non-local, unconstrained theory on (A)dS background, in the case of a
spin-$3$ field. To this end, we follow two independent procedures. 

 First, we look for a non-local compensator $\a_{\, \vf}$, in the spirit of  \cite{fms1}, 
and we make sure that the corresponding Lagrangian define the correct propagator. Then we consider
the functional integral for the \emph{local} theory, and we show that the integration of the compensator
$\a$ produces indeed the same effective, non-local Lagrangian. In this fashion, we stress once again,
the latter is given a clear interpretation, in terms of a theory whose dynamical content can be analysed 
using conventional methods.

\subsubsection{Construction of the non-local compensator} \label{sec3.3.1}

 If we were to follow the very same procedure we went through in \cite{fms1}, the first step would be 
to resort to some  proper, covariantised versions of the flat curvatures of \cite{dwf} (see e.g. \cite{manruhl}).
On the other hand, what we really need is a fully gauge invariant candidate ``Ricci'' tensor
on (A)dS, that in the spin-$3$ case can be more simply constructed 
resorting to the gauge transformation of the covariantised Fronsdal tensor, when no
assumptions are made on the traces of the gauge parameter\ft{In this way, looking for the kinetic tensor with the highest degree 
of singularity, we will be directly led to the solution possessing the correct propagator. If we were to start from curvatures, 
we would find proper (A)dS generalisations of the full family of Ricci-like tensors \eqref{Ricci3}. This simplification, however, 
is special of the spin-$3$ case.}. To begin with, it might be convenient to recall the deformed (A)dS Fronsdal operator
for a spin-$s$ field \cite{fronads}:
\be \label{fronads}
\cF_L \, = \, \cF \, - \, \frac{1}{L^{\, 2}} \, \left\{ \left[
(3\, -\, D\, -\, s)\, (2\, -\, s) \, - \, s \right]\, \vf \, + \, 2 \, g \, \vf^{\, \pe}
\right\} \, ,  
\ee

where $D$ denotes the space-time dimension and
\be {\cal F} \, = \, \Box \, \varphi \, - \, \nabla \, \nabla \cdot
\varphi \, + \, \nabla^{\, 2} \, \vf^{\, \pe}\,  \ee
is the (A)dS-covariantized Fronsdal operator. The  unconstrained variation of \eqref{fronads}
under  $ \delta \vf \, = \, \nabla \Lambda $, is
\be  \label{varAdS}
\d \cF_L \, = \, 3 \, \nabla^{\, 3} \Lambda^{\, \pe} \, - \,
\frac{4}{L^2}\, g\,  \nabla \, \Lambda^{\, \pe}  \, .
 \ee
Thus, considering the variation of  $\nabla \cdot \cF_L^{\, \pe}$,
\be
\d \, \nabla \cdot \cF_L^{\, \pe} \, = \, 3 \, \Box \, \{\Box \, - \, 
\frac{2}{L^{\, 2}}\,(D \, + \, 1)\} \, \L^{\, \pe} \, .
\ee 
it is  relatively simple to identify a candidate non-local compensator 
\be
\a_{\vf,\, L} \, = \, \fr{1}{3 \, \Box \, \{\Box \, - \, 
\frac{2}{L^{\, 2}}\,(D \, + \, 1)\}} \, \nabla \cdot \cF_L^{\, \pe} \, ,
\ee
allowing to define a gauge invariant tensor in the form
\be \label{adsA}
\cA_{\vf,\, L} \, = \, \cF_L \, - \, \{ 3 \, \nabla^{\, 3}\, - \, 
\fr{4}{L^{\, 2}} \, g \, \nabla\} \, \a_{\vf,\, L} \, .
\ee
The Bianchi identity satisfied by  $\cA_{\vf,\, L}$,
\be
\nabla \cdot \cA_{\vf,\, L} \, - \, \, \12 \, \nabla \, \cA_{\vf,\, L}^{\, \pe} \, \equiv \, 0 \, ,
\ee
together with its further property
\be
\nabla \cdot \cA_{\vf,\, L}^{\, \pe} \, \equiv \, 0 \, ,
\ee
easily verified from  
\be
\cA_{\vf,\, L}^{\, \pe} \, = \, \cF^{\, \pe}_L \, - \, 
\{3\, \nabla \, \Box \, - \, \fr{6}{L^{\, 2}} \, (D\, + \, 1) \, \nabla \} \, \a_{\vf,\, L} \, ,
\ee
imply the existence of a non-local, gauge-invariant  Lagrangian of the form
\be\label{nonlocAdS}
\cL \, = \, \12 \, \vf \, \{\cA_{\vf,\, L} \, - \, \12 \, g \, \cA_{\vf,\, L}^{\, \pe}\} \, - \, \vf \cdot \cJ \, .
\ee
It is then straightforward to evaluate the propagator, once the field is coupled 
to a conserved source. Indeed, following the procedure described in \cite{fms1, fms2}, from
the Lagrangian equation
\be
\cA_{\vf,\, L} \, - \, \12 \, g \, \cA_{\vf,\, L}^{\, \pe} \, = \, \cJ \, ,
\ee
we obtain
\be
\cA_{\vf,\, L} \, = \, \cJ \, - \, \fr{1}{D} \, g \, \cJ^{\, \pe} \, .
\ee
Then, introducing the Lichnerowicz operator $\Box_{\, L}$ defined by
\be 
\Box_L \, \varphi \, = \, \Box \, \varphi\ + {1\over {L^2}}\,
\left[s \, (D + s - 2)\, \varphi \, - \, 2 \, g \, \varphi^{\, \pe} \right] \, , 
\ee
we can rewrite $\cA_{\vf,\, L} $ for spin $3$ as
\be
\cA_{\vf,\, L} \, = \, (\Box_{\, L} \, - \, \fr{4}{L^{\, 2}}) \, \vf \, + \, 
\nabla \, (\nabla \cdot  \vf \, - \, \12 \, \nabla \, \vf^{\, \pe}) \,
- \,  (3 \, \nabla^{\, 3}\, - \, 
\fr{4}{L^{\, 2}} \, g \, \nabla) \, \a_L \, .
\ee
From this expression we can compute the interaction between conserved currents, finding
\be \label{lessprop}
\cJ \cdot \vf \, = \, \cJ \, \fr{1}{\Box_{\, L} \, - \, \fr{4}{L^{\, 2}}} \, \cJ \, 
- \, \fr{3}{D} \, \cJ^{\, \pe} \, \fr{1}{\Box_{\, L} \, - \, \fr{4}{L^{\, 2}}} \, \cJ^{\, \pe} \, ,
\ee
in agreement with the result found in \cite{fms1} for the local counterpart of \eqref{nonlocAdS}.

\subsubsection{Integration of the compensator from the local theory} \label{sec3.3.2}

 While the computation of the propagator already represents a strong consistency  check
for the  Lagrangian \eqref{nonlocAdS}, we can further clarify its interpretation if we can show
that it defines the effective Lagrangian of a more conventional local theory, 
once the compensator $\a$ is integrated away from the functional integral. 
To this end, following \cite{fms1}, let us write explicitly, for spin $3$, the extension to (A)dS of the local unconstrained Lagrangian \eqref{boselagr}:
\be \label{boselagrAdS}
{\cL} \, = \, \frac{e}{2} \, \vf \, \{\cA_L \, - \,
\frac{1}{2} \ g\, \cA_L^{\, \pe} \} \, - \, \frac{3\, e}{4}  \, \a \, \nabla \cdot \cA_L^{\, \pe} \, ,
\ee
where  $e$  denotes the determinant of the vielbein, $g$ is the (A)dS metric, while the tensor $\cA_L$ has the same 
form as in \eqref{adsA}, with   $\a$  being in this case an independent Stueckelberg field, s.t. $\d \, \a \, = \, \L^{\, \pe}$. 
In order to consistently eliminate the auxiliary field from the Lagrangian, we can start from 
\be \label{Z3AdS}
\cZ \, [\cJ] \, = \, \cN \, \int  \cD \vf \, \cD \a \, 
e^{\, i \, \int  d^d x \, e \, \{ \12 \, \vf \, \cE_{\, \vf}^L \, + \, \fr{9}{4} \, \a \, 
\Box \, [\Box \, - \, \fr{2}{L^{\, 2}} \, (D + 1)] \, \a \, 
- \, \fr{3}{2}\, \a \, \nabla \cdot \cF_L^{\, \pe} \, - \, \vf \, \cdot \, \cJ\}}\, ,
\ee
where we defined
\be
\cE_{\, \vf}^L \, = \, \cF_L \,- \, \12 \, g \, \cF^{\, \pe}{}_L\, ,
\ee
and perform the gaussian integration over $\a$, obtaining the effective Lagrangian
\be
\cL_{\, eff}\, (\vf) \, = \, \12 \, \vf \, \cE^L_{\, \vf} \, - \, \fr{1}{4} \, 
 \nabla \cdot \cF_L^{\, \pe}\fr{1}{\Box \, [\Box \, - \, \fr{2}{L^{\, 2}} \, (D + 1)]}   \nabla \cdot \cF_L^{\, \pe}\, ,
\ee
which, in its turn, can be shown to coincide with \eqref{nonlocAdS}, up to total derivatives.

\section{Effective non-local Lagrangians for fermions}  \label{sec4}

 Now we would like to extend our considerations to symmetric fermions. We start recalling 
the construction of the corresponding minimal local Lagrangians \cite{fs3, fms1}, to then pass to the integration
of the auxiliary field for the case of spin $\fr{5}{2}$, followed by  the geometrical interpretation of the result.

\subsection{Minimal local theory} \label{sec4.1}

 The construction of local, unconstrained Lagrangians for symmetric spinor-tensors\ft{In the conventions here followed
$\g_0$ is antihermitian, while the $\g_i$, $i = 1, 2, 3$ are hermitian. Moreover
$\g_0\, \g^{\, \dagger}_{\, \m} \, \g_0 \, = \,  \g_{\, \m}$. An additional useful combinatorial rule is
$\g \, \cdot \, (\g \, \psi) =  (D \, + \, 2 \, s) \, \psi \, - \, \g \, \psisl$ .} $\psi$ of rank $s$ (and spin $s + \12$) closely resembles the corresponding one for bosons, 
here sketched in Section \ref{sec2.1}. 
Under  the transformation $\delta \, \psi =  \pr \, \e$ the unconstrained variation of the Fang-Fronsdal tensor \cite{fangfronsdal},
\be \label{fangfr}
{\cS} \, = \, i\, \left(\dsll \, \psi \, - \, \pr \, \psisl
\right) \, , \ee
is $\delta \, {\cS} \, = \, -  \, 2 \, i\, \pr^{\, 2} \esl \, $.
Thus, we can build from $\cS$ the {\it fully} gauge invariant operator
\be \label{W}
{\cW} \, \equiv \, S \, + \, 2 \, i\, \pr^2 \xi \, , 
\ee
where the rank-$(s-2)$ \emph{compensator} $\xi$ transforms as $ \delta \, \xi \, = \, \esl$.
The Bianchi identity for $\cW$,
\be  \label{bianchifermi}
\prd \cW \, - \, \frac{1}{2} \ \pr \, \cW^{\, \pe} \, - \,
\frac{1}{2} \, \dsll \, \, {\cWsl}  \, = \, i\, \pr^{\, 2} \{ \,
\psisl^{\, \pe} \, - \, 2 \, \prd \xi \, - \, \pr \, \xi^{\, \pe} -
\not{\!  \pr} \xisl \, \} \, , 
\ee
leads naturally to a second gauge-invariant spinor-tensor,
\be \cZ \, \equiv \, i\, \bigl\{ \psisl^{\, \prime} \, - \, 2 \,
\prd \xi \, - \, \pr \, \xi^{\, \prime} - \not{\!  \pr} \xisl\bigr\}
\, , \ee
directly related to the triple $\g$-trace constraint on the
fermionic gauge field $\psi$, absent in the Fang-Fronsdal
formulation. The minimal flat-space Lagrangians of \cite{fs3, fms1} can
then be recovered starting from the trial Lagrangians
\be \cL_0 \, = \, \12 \, \bar{\psi} \, \{ \cW \, - \,
\frac{1}{2} \, \eta \, \cW^{\, \pe} \, - \, \frac{1}{2} \, \g \,
\cWsl \} \ + \ h.c. \, , \ee
and compensating the remainders in their gauge transformations with
new terms involving the field $\xi$ and the tensor $\cZ$. The
complete Lagrangian is finally
\be \label{fermilagr}
\begin{split}
\cL = & \12 \, \bar{\psi} \, \{ \cW \,- \, \frac{1}{2} \, \g \, {\not \! \! \! \cW}\, - \,
\frac{1}{2} \, \eta \, \cW^{\, \pe}
\} \, - \, \frac{3}{4} \, {n \choose 3} \,  {\not \! \bar{\xi}} \, \prd
\cW^{\, \pe} \\
& +\, \frac{1}{2} \, {n \choose 2} \, \bar{\xi} \, \prd {\not \! \! \! \cW} \, + \,
\frac{3}{2} {n \choose 3} \bar{\lambda} \, \cZ \,  + \ h.c. \, ,
\end{split}
\ee
where the Lagrange multiplier $\lambda$ transforms according to 
$\d \l =  \prd \prd \epsilon$,
in order for $\cL$ to be gauge invariant. Lagrangians 
for constrained spinor-tensors of \emph{any} symmetry, together with their
unconstrained extensions, were first presented in the second of \cite{cfms}.

\subsection{Integrating out auxiliary fields: flat background} \label{sec4.2}

 We would like to perform the integration over the auxiliary fields, and investigate the form of the corresponding effective non-local Lagrangian.
For simplicity, we limit ourselves to the example of spin  $s = \fr{5}{2}$, in which case \eqref{fermilagr}
reduces to\ft{To avoid confusion with signs: the last term in \eqref{lagr5/2} is to be interpreted as $\bar{\cS}_{\m}\, \g^{\, \m}$, and not as
${\not \! \cS}^{\, \dagger} \, \g^0$.}
\be \label{lagr5/2}
\cL \, = \, i \, \bar{\psi} \, \cE_{\, \psi} \, + \, 2 \, i \, \bar{\xi} \, \Box  \dsll \, \xi \, 
+ \, i \, (\bar{\xi} \, \prd {\not \! \cS}\, - \, \prd {\not \!  \bar{\cS}} \, \xi)\, ,
\ee
where we defined
\be
\cE_{\, \psi} \, = \, \cS \, - \, \12 \, \g \, {\not \! \cS} \, - \, \12 \, \h \, \cS^{\, \pe} \, . 
\ee
Thus, considering the fermionic partition function in the presence of two grassmannian sources
$\bar{\th}$ and $\th$,
\be \label{Zferm}
\cZ \, [\bar{\th}, \, \th] \, = \, \cN \, \int  \cD \bar{\psi} \, \cD \psi \,
\cD \bar{\xi} \, \cD \xi \, e^{\, i  \int  d^d x \, \{ \cL \, + \, \bar{\th} \, \psi \, + \, \bar{\psi} \, \th \}}\, ,
\ee
and performing the fermionic gaussian integration over $\bar{\xi}$ and $\xi$ we obtain 
the non-local effective Lagrangian
\be \label{effectiveferm}
\cL_{eff}\, (\bar{\psi},\, \psi) \, = \, i \, \bar{\psi} \, \cE_{\, \psi} \, + \, i \, 
\prd {\not \!  \bar{\cS}} \, \fr{\dsll}{2 \, \Box^{\, 2}} \, \prd {\not \! \cS} \, ,
\ee
which is equivalent, up to total derivatives, to the form
\be \label{effectiveferm}
\cL_{eff}\, (\bar{\psi},\, \psi) \, = \, i \, \bar{\psi} \, \{\cW_{\, \psi} \, - \, 
\12 \, \g \, {\not \! \! \cW}_{\, \psi} \, - \, \12 \, \h \, \cW^{\, \pe}_{\, \psi}\} \, ,
\ee
with the non-local, Dirac-like, kinetic tensor $\cW_{\, \psi}$ defined as 
\be \label{kinfermi}
\cW_{\, \psi} \, = \, \cS \, - \, \pr^{\, 2} \, \fr{\dsll}{\Box^{\, 2}} \, \prd {\not \! \cS} \, ,
\ee
and satisfying 
\be \label{bianchinonloc}
\prd {\not \! \! \cW}_{\, \psi} \, \equiv \, 0 \, ,
\ee
which is relevant to stress in view of the following discussion.
Now we would like to make it explicit the content of \eqref{effectiveferm} in terms of 
curvatures, and compare these results with those obtained in \cite{fs1, fs2, dariomass}
on the geometry of fermionic theories.

\subsection{Geometric interpretation and propagator} \label{sec4.3}

 Non-local candidate Dirac-Rarita-Schwinger tensors  were first proposed in \cite{fs1, fs2},
in analogy with the corresponding bosonic quantities computed from the curvatures of \cite{dwf}.
The issue of fermionic geometry was then reconsidered in \cite{dariomass}; there it was shown that, 
even keeping to a minimum the degree of singularity of the non-local operators involved, in the fermionic
case infinitely many independent tensors can actually be constructed. 

In particular, for  the case of spin $s = \fr{5}{2}$ of interest in this section, 
starting from the fermionic curvature
\be \label{curvature5/2}
\cR_{\, \m \m, \, \n \n} \, = \, \pr^{\, 2}_{\, \m} \, \psi_{\n \, \n} \, - 
\fr{1}{2} \, \pr_{\, \m} \, \pr_{\, \n}\, \psi_{\m \, \n} \, + \,  
\pr^{\, 2}_{\, \n} \, \psi_{\m\, \m} \, ,
\ee
\emph{two} independent, gauge-invariant, kinetic tensors can be constructed:
\be
\begin{split}
& \fr{\dsll}{\Box}\, \cR_2^{\, \pe} \equiv \, D_2\, , \\
& \fr{1}{\Box}\, \prd {\not \! \! \cR}   \equiv \, \hat{D}_2\, 
\end{split}
\ee
whose expression in terms of the Fang-Fronsdal operator \eqref{fangfr} is
\be \label{52second}
\begin{split}
&i\, D_2 \, =\, \cS \, + \,  \fr{\pr^{\, 2}}{\Box}\, 
\cS^{\, \pe} \, - \, \fr{\pr}{\Box}\, \prd \cS \, , \\
& i\, \hat{D}_2 \, = \, \cS \, - \, \12 \, \fr{\pr}{\Box}\, \prd \cS \, , \\
\end{split}
\ee
while the corresponding candidate Dirac tensor proposed in \cite{fs1, fs2},
\be
\cS_2 \, = \, \cS \, + \,  \fr{1}{3}\, \fr{\pr^{\, 2}}{\Box}\, \cS^{\, \pe} \, - 
\,\fr{2}{3}\,  \fr{\pr}{\Box}\, \prd \cS \, ,
\ee
can be shown to be equivalent to the linear combination
\be
\cS_2 \, = \, \fr{1}{3} \, i\, D_2 \, + \, \fr{2}{3} \, i\, \hat{D}_2  \, .
\ee
Moreover, in analogy with what already recalled for bosons in the previous sections, 
additional gauge invariant tensors can be constructed allowing for higher degrees
of singularity in the corresponding non-local expressions. On the other hand, the analysis of 
the current exchange performed for bosons in \cite{fms1} leads to expect that only one
of these theories would display the correct propagator. 

For spin $\fr{5}{2}$, in the  local formulation of \cite{fs3, fms1},
the field equation for $\psi$ in the presence of a source takes the form
\be \label{covfermiJ}
 {\cal W}\, - \, {1\over 2}\, \gamma\, \cWsl \, - \, {1\over 2}\ \eta\ {\cal W}^{\, \prime}\, = \, {\cal J}\, ,  
\ee
where the gauge invariant spinor-tensor $\cW$, defined in  \eqref{W}, satisfies the Bianchi identity
\eqref{bianchifermi} (where it is to be recalled that $\cZ = 0$, due to the equation for $\bar{\l}$), implying in particular that, on-shell
\be \label{bianchiloc}
\prd  \cWsl \, = \, 0\, ,
\ee
thus accounting for the conservation of $\cJ$.
Therefore, in order to reproduce the correct propagator, we should select a non-local Dirac tensor 
satisfying the same relations \eqref{bianchifermi} and \eqref{bianchiloc}, as well as an equation of the same form
as \eqref{covfermiJ}.

 It is not difficult to check that the Lagrangian theories associated to  Dirac-type kinetic tensors
\eqref{52second} (as well as any of their linear combinations) do \emph{not} reproduce
\eqref{covfermiJ}  nor \eqref{bianchiloc} . Differently, the non-local theory obtained after integration
of the compensators $\bar{\xi}$ and $\xi$, possesses exactly the correct properties, as
manifest from \eqref{effectiveferm} and \eqref{bianchinonloc}. Moreover, while other, independent, 
non-local compensators could be constructed, such as 
\be
\hat{\xi}_{\, \psi} \ = \, -\12 \, \fr{1}{i\, \Box} \, \cS^{\, \pe}\, ,
\ee
having the same gauge transformation as $\xi_{\, \psi} =  - \fr{\not \pr}{2\, i \, \Box^{\, 2}} \, \prd {\not \! \cS} $, nonetheless the corresponding kinetic tensor
\be
\hat{\cW}_{\, \psi} \, = \, \cS \, - \, \fr{\pr^{\, 2}}{\Box} \, \cS^{\, \pe} \, 
\ee
would \emph{not} satisfy  \eqref{bianchifermi} and \eqref{bianchiloc}.
This is another manifestation of the fact that, in the non-local case, it is actually possible to modify
quantities by the addition of non-local gauge invariant tensors in such a way that, while gauge transformation 
properties are obviously preserved, the meaning of the whole construction gets modified in a crucial way. 
Indeed, there is \emph{only one} geometric theory, for fermions of spin $\fr{5}{2}$, 
which propagates the correct number of degrees of freedom, and the procedure we followed 
in this paper can be seen as a safe way to derive it. Finally, its geometric interpretation
is made manifest expressing the kinetic tensor \eqref{kinfermi}
in terms of the curvature \eqref{curvature5/2}, according to
\be \label{fermigeom}
\cW_{\, \psi} \, = \, \cS \, - \, \pr^{\, 2} \, \fr{\dsll}{\Box^{\, 2}} \, \prd {\not \! \cS} \, 
= \, 2 \,  i\, \fr{1}{\Box}\, \prd {\not \! \! \cR}  \, - \,  i\, \fr{\dsll}{\Box}\, \cR^{\, \pe} \, + \,i \,  
\fr{\pr^{\, 2}}{\Box^{\, 2}} \, \dsll\, \cR^{\, \pe \pe}  \, .
\ee

\section{Summary and outlook} \label{sec5}

 We computed effective, non-local Lagrangians for some specific examples of higher-spin gauge fields
on flat and (A)dS backgrounds, performing the integration over non-physical fields in the corresponding local theories. 
The main goal of our computations has been to produce a \emph{definition} of non-local, 
geometric Lagrangians, possibly  devoid of ambiguities on their physical content, clarifying in particular the rationale behind the presence 
of  inverse powers of the d'Alembertian operator in their kinetic tensors.

  In this fashion, we were able to provide further support for the particular forms of geometric Lagrangians for symmetric 
bosons on flat backgrounds first given in \cite{fms1}. Here we also proposed  examples of 
geometric theories with correct propagators for fermions of spin $\fr{5}{2}$ on flat space-time, and 
for bosons of spin $3$ on (A)dS backgrounds. These results point towards the possibility that a geometric description 
of interacting higher-spins, involving proper deformations of  linearised curvature tensors, could emerge from a conventional, 
local theory, once all auxiliary fields (or at least a proper subset of them) are integrated away.  In particular, 
in view of the geometrical interpretation of triplets to be given elsewhere \cite{dariotripl}, it might not be necessary to this
end that the local theory be formulated in terms of \emph{unconstrained} fields.

 As for what concerns the latter, we leave for future work the generalisation of the results  here presented to 
irreducible, unconstrained bosons and fermions  of any spin (and symmetry), as well as the corresponding 
analysis of massive higher-spin fields.

\ack
I would like to thank the Organizers of the 1st Mediterranean Conference on Classical and Quantum Gravity for 
their kind invitation to present my lecture. 
I am grateful to A. Campoleoni, J. Mourad and  A. Sagnotti for useful exchanges and for collaboration on several topics 
reconsidered in this paper. I also wish  to thank the Scuola Normale Superiore of Pisa and the Institute of Physics 
of the ASCR in Prague for their hospitality while part of this work was being prepared. The present research 
was supported  by APC-Paris VII and  by the CNRS through the P2I program, and also in part by  the MIUR-PRIN contract 2007-5ATT78 and by the EURYI grant EYI/07/E010 from EUROHORC and ESF.

\section*{References}

\end{document}